\documentclass[10pt,conference]{IEEEtran}

\usepackage{cite}
\usepackage[dvips]{color}
\usepackage{tabu}
\usepackage{comment}
\usepackage{todonotes}
\usepackage{epsf}
\usepackage{epsfig}
\usepackage{epsfig}
\usepackage{graphicx}
\usepackage{bbold}
\usepackage{mathtools}
\usepackage{mathrsfs}
\usepackage{amssymb,amsmath}
\usepackage{pdfpages}
\usepackage{epstopdf}
\usepackage{algorithmicx}  
\usepackage[algo2e]{algorithm2e} 
\usepackage{dsfont}
\usepackage{lettrine} 
\usepackage{amsmath,epsfig,amssymb,algorithm,algpseudocode,amsthm,url}
\usepackage[justification=centering]{caption}
\usepackage{subcaption}
\usepackage{bbm}
\usepackage[utf8]{inputenc}
\usepackage{csquotes}
\usepackage{amsmath,amssymb}
\usepackage{verbatim}
\usepackage{amsmath,amssymb}
\usepackage{verbatim} 
\usepackage{setspace}	
\usepackage[T1]{fontenc}
\usepackage{textcomp}
\usepackage{amsmath}
\theoremstyle{definition}
\newtheorem{prop}{Proposition}
\usepackage{comment}
\usepackage{tabularx}
\usepackage{graphicx}
\usepackage{xpatch}
\DeclareMathOperator*{\argmax}{argmax}

\usepackage{stackengine}
\usepackage[nocomma]{optidef}
\def\delequal{\mathrel{\ensurestackMath{\stackon[1pt]{=}{\scriptstyle\Delta}}}}
\IEEEoverridecommandlockouts
\begin{document}
\title{\huge Performance Analysis and Optimization of Uplink Cellular Networks with Flexible Frame Structure 
\thanks{This research was supported by the U.S. National Science Foundation under Grants CNS-1941348 and CNS-2008646.} } \vspace{-0.5cm}
\author{
\authorblockN{Fatemeh Lotfi and Omid Semiari  }\\ \vspace*{-1em}
\authorblockA{\small Department of Electrical and Computer Engineering, University of Colorado, Colorado Springs, CO, USA\\ 
		Email: \protect\url{{flotfi,osemiari}@uccs.edu}\vspace{-0.5cm}\\
}}
\maketitle
\begin{abstract}
Future wireless cellular networks must support both enhanced mobile broadband (eMBB) and ultra-reliable low-latency communications (URLLC) to manage heterogeneous data traffic for emerging wireless services. To achieve this goal, a promising technique is to enable flexible frame structure by dynamically changing the data frame's numerology according to the channel information as well as traffic quality-of-service requirements. However, due to non-orthogonal subcarriers, this technique can result in an interference, known as inter numerology interference (INI), thus, degrading the network performance. In this work, a novel framework is proposed to analyze the INI in the uplink cellular communications. In particular, a closed-form expression is derived for the INI power in the uplink with a flexible frame structure and a new resource allocation problem is formulated to maximize the network spectral efficiency (SE) by jointly optimizing the power allocation and numerology selection in a multi-user uplink scenario. The simulation results validate the derived theoretical INI analyses and provide guidelines for the power allocation and numerology selection.  
\vspace{-0cm}
\end{abstract}
\section{Introduction} \vspace{-0cm}
Next-generation wireless networks are required to support new applications, such as connected and autonomous vehicles or wireless extended reality, that rely on both enhanced mobile broadband (eMBB) and ultra-reliable low-latency communications (URLLC) \cite{Omid2019}. To meet such strict service requirements, the 3rd generation partnership project (3GPP) has proposed the notion of flexible frame structure that allows dynamic selection of the numerology (i.e., sub-carrier spacing and cyclic prefix (CP) length) across successive data frame transmissions. Despite their advantages \cite{3gpp_st}, multi-numerology systems do not guarantee the subcarriers' orthogonality, resulting in the so-called inter numerology interference (INI).
Without proper interference mitigation, the INI can substantially impact the performance of URLLC services such as vehicular communications.\\
\indent The performance of multi-numerology systems and the impact of INI have been recently studied for downlink scenarios \cite{Mao2020,Kihero2019,syamfOFDM2020,akhtar2018,Marij2018,Liu2020,cheng2019,abu2018,Bag2019}. In particular, the authors in \cite{Mao2020} investigate the INI in the context of filtered orthogonal frequency division multiplexing (F-OFDM). In \cite{Kihero2019} the authors derive a closed-form expression for the INI in the downlink. The work in \cite{syamfOFDM2020} presents INI analysis for the windowed-OFDM (W-OFDM) scheme. Both works in \cite{Mao2020,syamfOFDM2020} use filtering and windowing techniques to control the out-of-band emission.
The authors in \cite{Bag2019} propose a new scheme that reduces the scheduling latency by leveraging standard numerology in conjunction with mini slots. An adaptive quality of service (QoS) aware numerology selection scheme is presented in \cite{akhtar2018} that aims to reduce the impact of INI in the downlink. In \cite{Marij2018}, the authors formulate the cost function based on the long-term average SINR of a multi-numerology system and propose a new resource allocation scheme that accounts for the Doppler and delay spreads. The authors in \cite{Liu2020} proposed an analytical expression of peak-to-average power ratio (PAPR) distribution for a mixed-numerology system and \cite{gok2020} presents two PAPR mitigation schemes for a multi-numerology cyclic prefix OFDM (CP-OFDM) system.
In \cite{cheng2019}, the authors present a precoding scheme to mitigate INI in the downlink and, \cite{abu2018} considers multi-numerology in non-orthogonal multiple access (NOMA) using successive interference cancellation. The authors in \cite{cheng2019UL} present an INI cancellation method for the uplink based on the minimum mean square error (MMSE) detection in OFDM systems. In \cite{yang2020}, a radio access network slicing framework is proposed, that by considering both radio frequency and baseband imparities, defines numerology relationship among slices.
While interesting, the works in \cite{Mao2020,Kihero2019,syamfOFDM2020,akhtar2018,Marij2018,Liu2020,cheng2019,abu2018}, only consider the INI analysis in the downlink, while as we show here, the INI can have a major impact on the uplink performance. 
Also, the body of work in \cite{Mao2020,Kihero2019,syamfOFDM2020,akhtar2018,Marij2018,Liu2020,cheng2019,abu2018,yang2020,Bag2019,cheng2019UL,gok2020} does not study the uplink performance for Discrete Fourier Transform-spread OFDM (DFT-s-OFDM) which is the common modulation scheme for uplink transmissions \cite{23GPP38}. Compared to OFDM, the DFT-s-OFDM transmission results in a lower PAPR for uplink signals and less power consumption at the user equipment (UE).\\
\indent The main contribution of this paper is a novel scheme for analysis and optimization of spectral efficiency (SE) in uplink cellular networks with flexible frame structure. First, we derive a closed-form expression for the INI in the multi-numerology system with the uplink DFT-s-OFDM transmission scheme. \textit{To the best of our knowledge, this is the first work that derives the INI for uplink DFT-s-OFDM networks}.
Building on this analysis, we formulate a new optimization problem that aims to maximize the minimum SE across the network while accounting for the impact of INI. Then, to solve this problem, we propose a new solution that yields the subcarrier powers allocation jointly with the numerology for each UE to optimize the network's performance. Comprehensive simulation results are presented to validate the derived INI analysis and to provide guidelines for joint power allocation and numerology selection in uplink cellular networks.
The paper is organized as follows. Section II describes the system model. Section III presents the INI analysis in the multi-user uplink scenario with multi-numerology support. Section IV presents the problem formulation and the proposed solution to optimize the SE. Simulation results are presented in Section V. Section VI concludes the paper.
\vspace{-0cm}
\section{System Model}
\begin{figure}[t!]
   \centering
    \includegraphics[width=7.8cm]{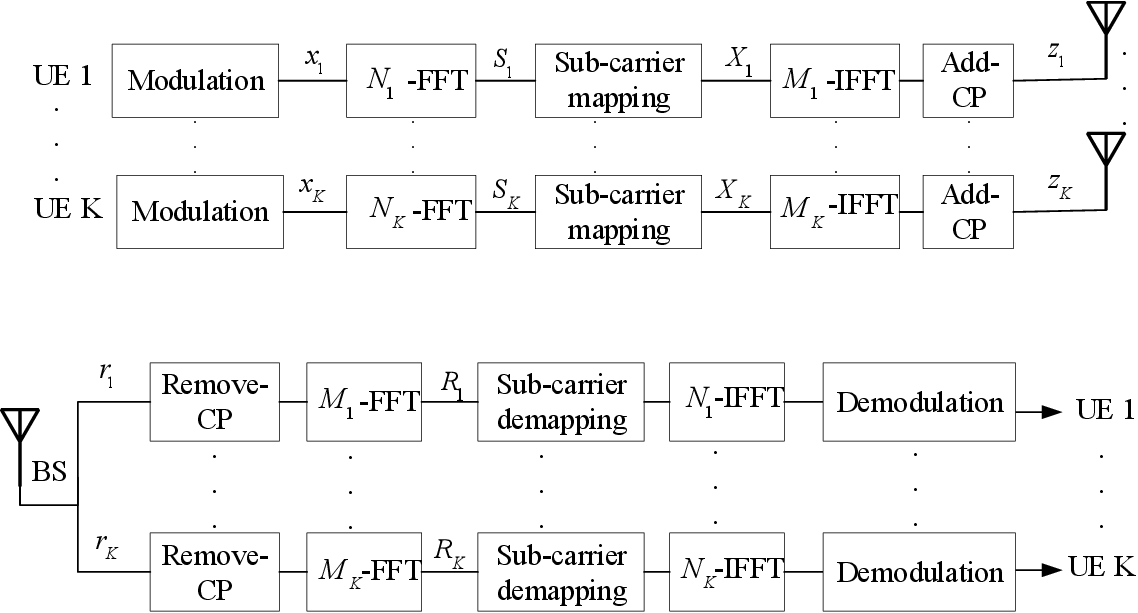}\vspace{-0cm}
    \caption{\small A block diagram of a DFT-s-OFDMA.}\vspace{-0.4cm}
    \label{sc-fdma}
\end{figure}

We consider an uplink multi-numerology cellular network consists of $K$ UEs in a set $\mathcal{K}$ and a base station (BS). The uplink signals are modulated based on the DFT-s-OFDM~\cite{3gpp_st} scheme, as shown in Fig. \ref{sc-fdma}. The uplink network supports a flexible frame structure which allows UEs to dynamically select different numerologies (i.e., subcarrier spacing and CP length)~\cite{3gpp_st}. 
At the transmitter of an arbitrary UE $i \in \mathcal{K}$, the modulated data stream at the output of the fast Fourier transform (FFT) block can be written as
\begin{align}
    S_i(k) = \sum_{n=0}^{N_i-1} x_i(n)e^{-j\frac{2\pi n k}{N_i}},
\end{align}
where the FFT size, $N_i$, is equal to the number of active subcarriers allocated to the $i$-th UE and $S_i(k)$ is the FFT symbol of the $k$-th subcarrier. Moreover, $x_i(n)$ is the modulated symbol of UE $i$. Note that $S_i(k)$, are independent and identically distributed (i.i.d) complex modulation symbols with zero mean and unit average power.
The numerologies of the UEs are determined as follow:
\vspace{-0.1cm}
\begin{align}
 &M_i=N_i+N_{z_i}, \\
 &\frac{N_1}{M_1}+\cdots+\frac{N_{K}}{M_{K}}=1, \vspace{-0.2cm}
\end{align}
where $N_{z_i}$, $M_i$, and $\frac{N_i}{M_i}$ are, respectively, the number of zero-padded subcarriers, the total number of subcarriers assigned to UE $i$, and the fraction of bandwidth allocated to the $i$-th UE. By zero-padding both the left and right sides of the signal subcarriers, the signal can be mapped into the determined position within the system's bandwidth.
To this end, a localized subcarrier mapping technique can be implemented whose output can be written as
\begin{align}\label{ScP1}
    X_i(k) = \begin{cases}
     S_i(k), & \text{$N_{i-1}+1\leqslant k \leqslant N_{i-1}+N_i-1$},\\
      0, & \text{otherwise}.    
    \end{cases}
\end{align}
\indent The signal after the inverse FFT (IFFT) block will be
\vspace{-0.1cm}
\begin{align}
    \Bar{x}_i(m) &= \frac{1}{M_i}\sum_{l=N_{i-1}+1}^{N_i-1} X_i(l)e^{j2\pi\frac{ m l}{M_i}},
\end{align}
where $1\leq m\leq M_i$. Finally, a cyclic prefix ($\tau_{\text{cp,i}}$), equal to $M_i/16$, is appended to the UE $i$ symbols before concatenating symbols. 
Let $\alpha_{i,j} = \frac{M_i}{M_j}$ be the \textit{numerology scaling factor}, where $i\neq j$, for $i,j \in \mathcal{K}$. For an arbitrary UE $i$, we can partition the user set $\mathcal{K}$ into three disjoint subsets, based on the values of numerology scaling factor. We define $\mathcal{K}_{h,i}=\{j\in\mathcal{K}\mid \alpha_{i,j}>1\}$ as the subset of UEs with higher numerology compared to UE $i$, $\mathcal{K}_{l,i}=\{j\in\mathcal{K}\mid \alpha_{i,j}<1\}$ as a subset of UEs with lower numerology and $\mathcal{K}_{s,i}=\{j\in\mathcal{K}\mid \alpha_{i,j}=1\}$ as a subset of UEs with the same numerology as UE $i$. Next, we will use the defined subsets to analyze the INI.\\
\indent Considering different numerology of UEs $i$ and $j$ and $\alpha_{i,j}$, the symbol duration of UE $i$ is equal to $\alpha_{i,j}$ symbol duration of UE $j$. So, in one symbol duration of UE $i$ the transmitted signals of UE $i$ and UE $j$ can be written as  
\vspace{-0cm}
\begin{equation}\label{y}
    z_i(m) =\Bar{x}_i(m),\,\,\,\, z_j(m) = \sum_{q=0}^{\alpha_{i,j}-1}\Bar{x}_j(m-\nu_q),\\
\end{equation}
where $\nu_q = q(M_j+\tau_{\text{cp,j}})$. Next, we represent the received signals at the BS under the flexible numerology setting.
Let $h_i$ represents the channel impulse response for UE $i$'s link. As shown in Fig. \ref{sc-fdma}, the received signal of UE $i \in \mathcal{K}$ for one symbol duration can be written as 
\begin{equation}
\begin{aligned}
    r_i(m) &= z_i(m) \ast h_i(m)+ \bigg(\sum_{j\in\mathcal{K}_{h,i}}^{}z_j(m) \ast h_j(m)\\& + \sum_{j\in \mathcal{K}_{l,i}}^{}z_j(m) \ast h_j(m)\bigg)+ w(m),\\
    \end{aligned} \label{rm}
\end{equation}
where $z_i(m)$ is the desired transmitted symbol of UE $i$, and $w(m)$ is the additive white Gaussian noise (AWGN) with zero mean and variance $\sigma^2$. In \eqref{rm}, $z_j(m)$ is the $\alpha_{i,j}$ symbols of UE $j$, that have been aligned with one symbol of the arbitrary UE $i$. Hence, the second term in \eqref{rm} will results in INI, as shown in the next section.
\section{Uplink Inter Numerology Interference Analysis}
At the BS, the received signal for an arbitrary UE $i$ after CP removal and passing through the FFT block can be written as
\begin{equation}\label{s_r1}
\begin{aligned}
&R_{i}(k)=\sum_{m=0}^{M_i-1}r_i(m)e^{-j2\pi\frac{ m k}{M_i}}=\sum_{m=\tau_{\text{cp,i}}}^{M_i-1}\Big(\Bar{x}_i(m)\ast h_i(m)+\\&\sum_{j\in \mathcal{K}_{h,i}}^{} \sum_{q=0}^{\alpha_{i,j}}
\Bar{x}_j(m-\nu_q)\ast h_j(m,q)+\sum_{j\in \mathcal{K}_{l,i}}^{}\Bar{x}_i(m)\ast h_i(m,q)\\&+w(m)\Big)e^{-j2\pi\frac{ m k}{M_i}}\delequal{R_{d_i}(k) + R_{I_{h,i}}(k)+ R_{I_{l,i}}(k) + W(k)}.
\end{aligned}
\end{equation}
\indent In \eqref{s_r1}, the first term is the desired signal. However, the second and third terms in \eqref{s_r1}, $R_{I_{h,i}}$ and $R_{I_{l,i}}$, represent, respectively, the INI caused by UEs in $\mathcal{K}_{h,i}$ and $\mathcal{K}_{l,i}$. We note that there will be no INI from the UEs in $\mathcal{K}_{s,i}$ since they have the same numerology as UE $i$ and their transmissions will be orthogonal to UE $i$'s link.  From \eqref{s_r1}, we can derive the following results for the INI in the uplink. 
\vspace{-0cm}
\begin{prop} 
For a UE $i \in \mathcal{K}$, the received INI at subcarrier $n$ from other UEs with higher and lower numerology (compared to UE $i$) in the uplink can be written, respectively, as   
\vspace{-0.4cm}
\begin{align} 
&I_{h,i}(n) = \frac{1}{N_i}\sum_{k=0}^{N_i-1}\sum_{j\in\mathcal{K}_{h,i}}^{}\sum_{q=1}^{\alpha_{i,j}} \sum_{l=0}^{N_j-1} \frac{1}{M_j}X_j(\nu_l)H_j(\nu_l) \,\times \nonumber\\ &\hspace{3.5cm}Q_{i,j,h}^{I}(l,k,q)
{\rm e}^{-j2\pi(\frac{\nu_{q-1}}{M_i}-\frac{n}{N_i})},\label{I1}\\
&I_{l,i}(n) = \frac{1}{N_i}\sum_{k=M_i-N_i}^{M_i-1}\sum_{j\in\mathcal{K}_{l,i}}^{}\sum_{q=1}^{\frac{1}{\alpha_{i,j}}}\sum_{l=0}^{N_j-1}\frac{1}{M_j} X_j(l)H_j(l)\,\times\nonumber\\
&\hspace{3.6cm}Q_{i,j,l}^{I}(l,k,q) 
{\rm e}^{j2\pi\frac{n k}{N_i}},\label{I2}
\end{align}
\vspace{-0cm}
where $\tau_{\text{cp},j}$ is the CP length of UE $j$  and
\begin{align}
    &\nu_l = l+M_j-N_j, \hspace{0.5cm} \nu_q = q(M_j+\tau_{\text{cp},j}),\\
    &\kappa_{l,k} = \frac{l}{M_j}+\frac{N_i-k}{M_i},\\
    &V_q = \begin{cases}
    M_j+(2-\alpha_{i,j})\tau_{\text{cp},j}, & q=1\\
    M_j+\tau_{\text{cp},j}, & q>1
    \end{cases}.
\end{align}
\indent In addition,
\begin{align}
    &Q_{i,j,h}^{I}(l,k,q) = {\rm e}^{j\pi\kappa_{l,k}(V_q-1)}\frac{\sin\big(\pi\kappa_{l,k}V_q\big)}{\sin\big(\pi\kappa_{l,k}\big)},\\
    &Q_{i,j,l}^{I}(l,k,q) = \begin{cases}
     Q_{l,1}^{I} {\rm e}^{j2\pi(M_j-\tau_{\alpha})\zeta_{l,k}} + Q_{l,2}^{I}, & \text{$q=1$}\\
     Q_{l,3}^{I} {\rm e}^{j2\pi\zeta_{l,k}(\nu_q-M_i)}, & \text{$q>1$}
    \end{cases},\\
    &Q_{l,1}^{I} = {\rm e}^{j\pi\zeta_{l,k}(\tau_{\alpha}-1)}\frac{\sin\big(\pi\zeta_{l,k}\tau_{\alpha}\big)}{\sin\big(\pi\zeta_{l,k}\big)},\\
    &Q_{l,2}^{I} = {\rm e}^{j\pi\zeta_{l,k}(M_i-\tau_{\alpha}-1)}\frac{\sin\big(\pi\zeta_{l,k}(M_i-\tau_{\alpha})\big)}{\sin\big(\pi\zeta_{l,k}\big)},\\
    &Q_{l,3}^{I} = {\rm e}^{j\pi\zeta_{l,k}(M_i-1)}\frac{\sin\big(\pi\zeta_{l,k}M_i\big)}{\sin\big(\pi\zeta_{l,k}\big)},\\
    &\zeta_{l,k} = \frac{l}{M_j}-\frac{k}{M_i},\hspace{0.5cm} \tau_{\alpha} = (\frac{1}{\alpha_{i,j}}-1)\tau_{\text{cp,i}}.
\end{align}
\end{prop}
\vspace{-0cm}
\begin{IEEEproof}
The proof is presented in Appendix A.
\end{IEEEproof}
\vspace{-0cm}
Using the results in Proposition 1, the total INI at the UE $i$'s link will be
\begin{equation}\label{I1m}
    I_i(n) = I_{h,i}(n) + I_{l,i}(n), 
\end{equation}
where $I_{h,i}(n)$ and $I_{l,i}(n)$ given in \eqref{I1} and \eqref{I2}, respectively. 
Next, using this analysis, we will formulate a new optimization problem to maximize the minimum SE across the uplink network while accounting for the INI.
\section{Joint Numerology Selection and Power Allocation in Uplink Cellular Communications }
\vspace{-0cm}
Given the derived INI, the SE for UE $i$'s link will be
\begin{align}\label{SE}
S_{e_i}(n) = (1-P_{\text{out}})\log_{2}\Big(1+\frac{p_{d,i}(n)}{p_{I,i}(n)+\sigma^2}\Big),
\end{align}
where $p_{d,i}(n)$ and $p_{I,i}(n)=|I_i(n)|^2$ are defined, respectively, as the received desired power and the INI power. Also, $P_{\text{out}}$ represents the outage probability.
To optimize the SE across the uplink network we define $\Lambda(\boldsymbol{p},\boldsymbol{M})$ as  
\begin{align}\label{Robj}
    \Lambda(\boldsymbol{p},\boldsymbol{M}) =\min(\boldsymbol{S_{e}}(n)),
\end{align}
where $\boldsymbol{S_{e}}$ is a vector of $S_{e_i}$, $i\in\mathcal{K}$. 
Also, $\boldsymbol{p}$ and $\boldsymbol{M}$ represent the vector of subcarriers' powers and the vector with elements $M_i$ with $i\in\mathcal{K}$, respectively. 
Our goal is to optimize the system performance by maximizing $\Lambda$, while taking into account the INI. To this end, we formulate the following problem that aims to maximize the minimum SE across the uplink network:
\begin{align}
\argmax_{\boldsymbol{p},\boldsymbol{M}} & \hspace{0.7cm} 
\Lambda\left({\boldsymbol{p}},{\boldsymbol{M}}\right),\label{opt1}\\
\text{s.t.,} 
& \hspace{0.7cm} {S}_{e_{i}}(n)\geq {\lambda}_{i},\,\,\, i \in \mathcal{K},\\
& \hspace{0.7cm} 0\leq {p}_{x_{i}}\leq {p}_{\max},\\
& \hspace{0.7cm} {M}_{i}=2^{\mu_{i}}, \hspace{0.2cm} \mu_{i}\in \{6,7,8,9,10\},\label{optM}
\end{align}
where $\lambda_{i}$ is the minimum required SE for data transmission, $p_{x_i}$ is the subcarriers' power of UEs $i$ and ${p}_{\max}$ is the maximum allowed power per subcarrier. Note that the objective function and the first constraint are non-convex. To solve this problem, our proposed solution is based on relaxing the non-convex constraint and solving the problem as a convex optimization problem.
The proposed solution can be described as follow. For a given $\boldsymbol{M}$, we first optimize the transmit power based on the successive convex optimization technique\cite{boyd_vandenberghe_2004}. Then, given the obtained transmit power, we find the optimal numerology factors in $\boldsymbol{M}$ by simply searching over the limited possible values presented in \eqref{optM}.\\
\indent To convert the problem into a convex optimization, note that in \eqref{SE}, the outage probability can be shown to be convex \cite{poutageconvex}. Moreover, the log part can be written by expanding it as a difference between two concave functions as follow:
\begin{align}
    S_{e_i}(n) =& (1-P_{\text{out}})\Big(\log_{2}(p_{I,i}(n)+\sigma^2+p_{d,i}(n))\nonumber\\
    &\hspace{4.5cm}-S_{e_{i},sub}\Big),\label{sei}
    \end{align}
    \vspace{-0.3cm}
where,
    \begin{align}
    S_{e_{i},sub} = \log_{2}\Big(p_{I,i}(n)+\sigma^2\Big).
\end{align}
Since in \eqref{sei} both the subtrahend ($\text{SE}_{i,sub}$) and minuend are concave, there is no guarantee that their difference is also concave. Here, we can define an upper bound for $S_{e_i,sub}$ based on its first-order Taylor approximation. By minimizing this upper bound, we can effectively maximize the SE \cite{boyd_vandenberghe_2004}. Thus, we derive the following convex upper bound at the given local point $\hat{p}_{x_{i,r}}$:
\begin{equation}\label{rate1_p_sub}
\begin{aligned}
   & S_{e_{i},sub} \leq\Big(D_{i}(p_{x_{i}}-\hat{p}_{x_{i,r}})+\Upsilon_i\Big)\delequal S_{e_{i},sub}^{up}, 
\end{aligned}
\end{equation}
\vspace{-0.1cm}
where
\begin{align}
    &\Upsilon_i=\{S_{e_{i},sub}\mid p_{x_{i}}=\hat{p}_{x_{i,r}}\},\\
   &D_{i,p} =\{\frac{\partial S_{e_{i},sub}}{\partial p_{x_{i}}}\mid p_{x_{i}}=\hat{p}_{x_{i,r}}\},
\end{align}
and $\hat{p}_{x_{i,r}}$ is the given transmit power in the $r$-th iteration of the convex optimization. As a result, with the upper bound in \eqref{rate1_p_sub}, problem \eqref{opt1} for any given $M_i$ parameters, is a convex optimization problem that can be solved efficiently by an existing convex optimization approach \cite{boyd_vandenberghe_2004}.\\
\vspace{-0.5cm}
\section{Numerical Results}
We consider a multi-numerology uplink system with three UEs where UE 1 uses subcarrier spacing of $\Delta f_1 = 60$  kHz with $M_1 = 256$ as its FFT order, while UE 2 and UE 3 may switch between other numerologies with $\alpha_{1,2} = \{2,4,8\}$ and $\alpha_{1,3} = \{2,4,8\}$. We let $\tau_{{\rm cp},1} = M_1/16$  and use quadrature amplitude modulation (QAM) with $64$ symbols and unity average power. We further consider Rayleigh fading channel and a maximum transmit power of $100$ mW for uplink transmissions.\\
\begin{figure}
   \centering
    \includegraphics[width=7.8cm]{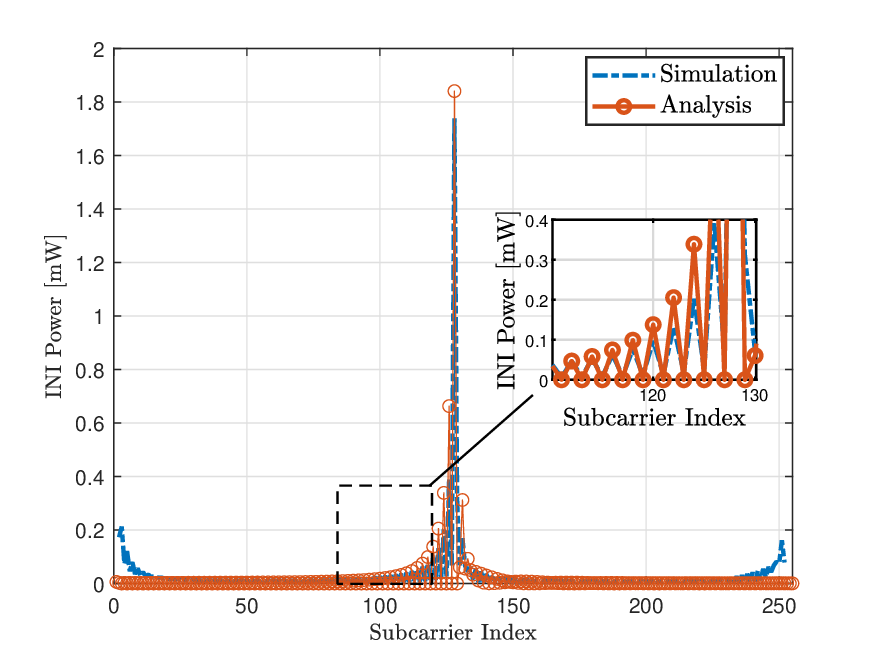}\vspace{-0.2cm} 
    \caption{\small INI power vs. subcarrier index with respect to the subcarrier spacing granularity of UE 1.}\vspace{-0.4cm}
    \label{INIpow}
\end{figure}
Fig. \ref{INIpow} compares the INI power per subcarrier obtained from simulations with the derived theoretical results for $\alpha_{1,2} = 2$. From Fig. \ref{INIpow}, we first note that the simulations validate the derived analyses in Proposition 1. We also observe that the edge subcarriers of both numerologies are more impacted by the INI, as compared to the middle subcarriers. The main reason behind this result is that the sinc-like signals corresponding to each subcarrier decays with a rate of $1/f$ \cite{Mao2020}. Moreover, the sinc functions of the two adjacent BWPs are not orthogonal at the sampling points anymore, due to using different numerologies in the uplink. Hence, the maximum level of INI is observed at the border of the two BWPs or edge subcarriers.\\ 
\begin{figure}
   \centering  
    \includegraphics[width=7.8cm]{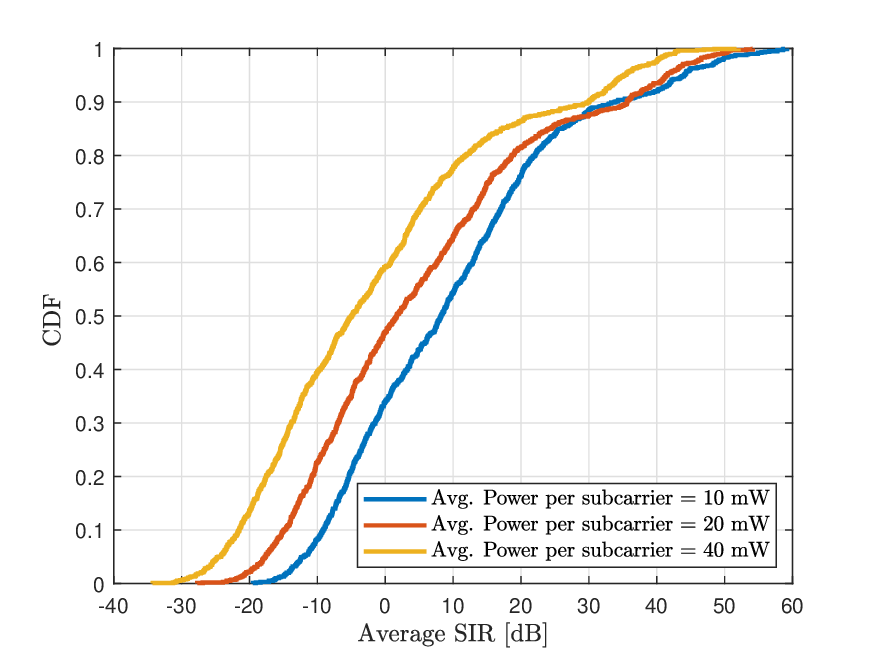}\vspace{-0.2cm}
    \caption{\small The CDF of average SIR for UE 1.}\vspace{-0.4cm}
    \label{cdf}
\end{figure}
Fig. \ref{cdf} shows the cumulative distribution function (CDF) of the signal to INI power ratio (SIR) for randomly selected $\alpha_{1,2}$ and $\alpha_{1,3}$ and different average power per subcarrier. Fig. \ref{cdf} shows that with random numerology selection, the SIR can be degraded while increasing the average power per subcarrier. Therefore, it is meaningful to control the INI by jointly optimizing the power allocation and the numerology selection. \\
\begin{figure}
   \centering 
    \includegraphics[width=7.8cm]{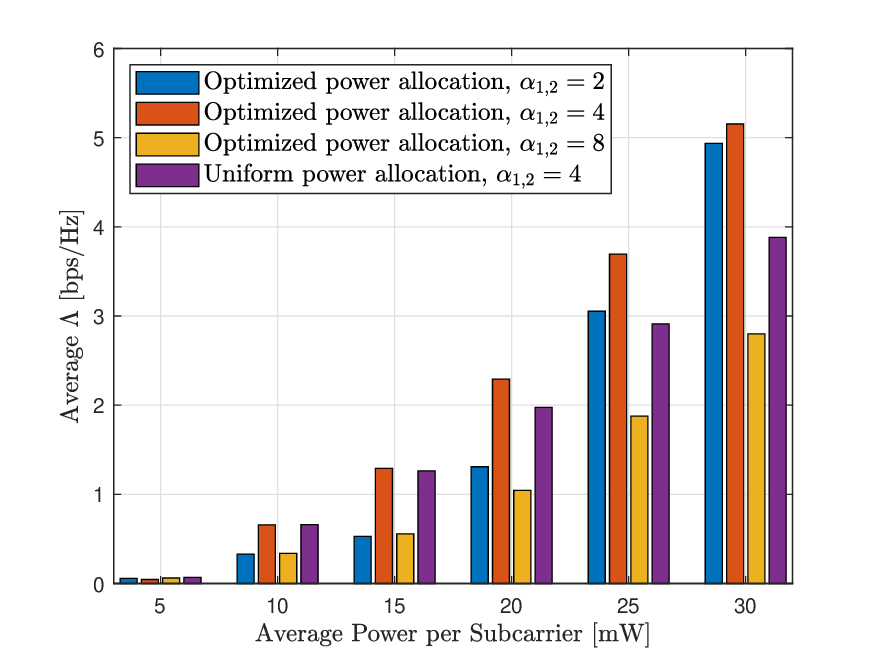}\vspace{-0.2cm}
    \caption{\small Average $\Lambda$ vs. the average power per subcarrier. }\vspace{-0.4cm}
    \label{avg_se_3UE}
\end{figure}
Fig. \ref{avg_se_3UE} shows the effect of multi-user interference on the average $\Lambda$ as the minimum SE across the UEs, with $\alpha_{1,2}=\{2,4,8\}$, $\alpha_{1,3} = 2$. The results are also compared with uniform power allocation with $\alpha_{1,2}=4$ and $\alpha_{1,3}=2$. The results show that for $\alpha_{1,2}=4$, the performance is better than $\alpha_{1,2}=\{2,8\}$, and thus, highest or lowest $\alpha$ does not necessarily yield the optimal performance. Fig. \ref{avg_se_3UE} also shows that the proposed solution can yield up to 30\% performance gain compared to uniform power allocation.
\vspace{-0cm}
\section{Conclusions}
In this paper, we have proposed a novel scheme to analyze and optimize the performance of uplink cellular communications with flexible frame structure. We have presented the theoretical derivations for the INI power in the uplink and have formulated a new resource allocation problem that maximizes the minimum SE across the network by jointly optimizing the power allocation and the numerology selection. The simulation results have corroborated the derived INI analyses and have provided guidelines for joint power allocation and numerology selection in uplink communications. \vspace{0cm}
\begin{appendices}
\section{proof of proposition 1}
\vspace{-0.2cm}
Let $X_i$ and $H_i$ represent the FFT of the transmitted signal and the channel frequency response of an arbitrary UE $i$, respectively. From \eqref{s_r1}, the interference term resulting from INI can be written as
    \begin{align}
        &R_{I_{h,i}}(n) =\sum_{j\in\mathcal{K}_{h,i}}^{} \sum_{m=\tau_{\text{cp,i}}}^{M_j+\tau_{\text{cp},j}-1}\frac{1}{M_j}\Big(\sum_{l=0}^{N_j-1}X_j(\nu_l)H_j(\nu_l)e^{\frac{j2\pi}{M_j} \nu_l m}\Big)\nonumber\\
        &e^{-\frac{j2\pi}{M_i}m k}+\sum_{j\in\mathcal{K}_{h,i}}^{}\sum_{q=2}^{\alpha_{i,j}}\sum_{m=\nu_{q-1}}^{\nu_q-1}\frac{1}{M_j}\Big(\sum_{l=0}^{N_j-1}X_j(\nu_l)e^{-\frac{j2\pi}{M_j}\nu_{q-1}\nu_l}\nonumber\\\label{sI1}
        &H_j(\nu_l)e^{\frac{j2\pi}{M_j} \nu_l m}\Big)e^{-\frac{j2\pi}{M_i}m k}.
    \end{align}
Similarly, for a neighbor UEs $j$ with lower numerology, the $I_{l,i}$ can be written as
    \begin{align}
        &R_{I_{l,i}}(k) = \sum_{j\in\mathcal{K}_{l,i}}^{}\sum_{m=M_j-\tau_{\alpha}}^{M_j-1}\frac{1}{M_j}\Big(\sum_{l=0}^{N_j-1}X_j(l)H_j(l)e^{-\frac{j2\pi}{M_j} ml}\Big)\nonumber \\
        &e^{-\frac{j2\pi}{M_i}m k}      
        +\sum_{j\in\mathcal{K}_{l,i}}^{}\sum_{m=0}^{M_i-\tau_{\alpha}-1}\frac{1}{M_j}\Big(\sum_{l=0}^{N_j-1}X_j(l)H_j(l)e^{-\frac{j2\pi}{M_j} ml}\Big)\nonumber\\
        &e^{-\frac{j2\pi}{M_i}m k}
        +\sum_{j\in\mathcal{K}_{l,i}}^{}\sum_{q=2}^{\frac{1}{\alpha_{i,j}}} \sum_{m=\nu_q-M_i}^{\nu_q-1} \frac{1}{M_j}\Big(\sum_{l=0}^{N_j-1}X_j(l)H_j(l)\nonumber\\\label{sI2}
        &e^{-\frac{j2\pi}{M_j} ml}\Big)
        e^{-\frac{j2\pi}{M_i}m k}.
    \end{align}
Then, the received signal at subcarrier $n$ after subcarrier demapping and IFFT can be written as 
  \begin{align}
    &\Tilde{R_i}(n)=\frac{1}{N_i}\sum_{k=0}^{N_i-1}\Big(R_{d_i}(k) + R_{I_i}(k) + W(k) \Big)e^{j\frac{2\pi}{N_i} kn},\nonumber\\\label{r_12}
    & \delequal{d_i(n) + I_i(n) + \Tilde{w}(n)}.
\end{align}
The INI term from \eqref{sI1} and \eqref{sI2} can be written as\\ \vspace{-0cm}
\begin{equation}
 I_{h,i}(n) =
 \frac{1}{N_i}\sum_{k=0}^{N_i-1}\sum_{j\in\mathcal{K}_{h,i}}^{}\frac{1}{M_j}\Bigg(\sum_{m=\tau_{\text{cp,i}}}^{M_j+\tau_{\text{cp,j}}-1}\Big(\sum_{l=0}^{N_j-1}X_j(\nu_l)H_j(\nu_l) \nonumber
 \end{equation}
 \begin{align}
 e^{\frac{j2\pi}{M_j} \nu_l m}\Big)e^{-\frac{j2\pi}{M_i}m k}+\sum_{q=2}^{\alpha_{i,j}}\sum_{m=\nu_{q-1}}^{\nu_q-1}\Big(\sum_{l=0}^{N_j-1}X_j(\nu_l)e^{-\frac{j2\pi}{M_j}\nu_{q-1}\nu_l} \nonumber
  \end{align}
\begin{align}
H_j(\nu_l)e^{\frac{j2\pi}{M_j} \nu_l m}\Big)e^{-\frac{j2\pi}{M_i}m k}\Bigg)e^{\frac{j 2\pi}{N_i} n k}=\frac{1}{N_i}\sum_{k=0}^{N_i-1}\sum_{j\in\mathcal{K}_{h,i}}^{}\sum_{q=1}^{\alpha_{i,j}} \sum_{l=0}^{N_j-1}\nonumber
\end{align}
\begin{align}\label{tsI1}
 \frac{1}{M_j}X_j(\nu_l)H_j(\nu_l)e^{-j2\pi(\frac{\nu_{q-1}}{M_i}-\frac{n}{N_i})}Q_{i,j,h}^{I}(l,k,q).
\end{align}
In \eqref{r_12}, the lower and upper limits of the IFFT summation are defined according to \eqref{ScP1}. In the following, the received INI signal from neighbor UE $j$ with lower numerology can be written as
\begin{align}
        I_{l,i}(n) =\frac{1}{N_i}\sum_{k=0}^{N_i-1}\sum_{j\in\mathcal{K}_{l,i}}^{}\frac{1}{M_j}\Bigg(\sum_{m=M_j-\tau_{\alpha}}^{M_j-1}\Big(\sum_{l=0}^{N_j-1}X_j(l)H_j(l)\nonumber
\end{align}
\begin{align}
        e^{-\frac{j2\pi}{M_j} ml}\Big)e^{-\frac{j2\pi}{M_i}m k}+\sum_{m=0}^{M_i-\tau_{\alpha}-1}\Big(\sum_{l=0}^{N_j-1}X_j(l)H_j(l)e^{-\frac{j2\pi}{M_j} ml}\Big)\nonumber
\end{align}
\begin{align}
        e^{-\frac{j2\pi}{M_i}m k}
        + \sum_{q=2}^{\frac{1}{\alpha_{i,j}}} \sum_{m=\nu_q-M_i}^{\nu_q-1}\Big(\sum_{l=0}^{N_j-1}X_j(l)H_j(l)e^{-\frac{j2\pi}{M_j} ml}\Big)\nonumber
\end{align}
\begin{align}
       & e^{-\frac{j2\pi}{M_i}m k}\Bigg)e^{\frac{j2\pi}{N_i}n k}
        =\frac{1}{N_i}\sum_{k=M_i-N_i}^{M_i-1}\sum_{j\in\mathcal{K}_{l,i}}^{}\sum_{q=1}^{\frac{1}{\alpha_{i,j}}}\sum_{l=0}^{N_j-1}\frac{1}{M_j}X_j(l)\\ \nonumber
       & H_j(l)Q_{i,j,l}^{I}(l,k,q)e^{\frac{j2\pi}{N_i}n k}.\label{tsI2}
\end{align}
\vspace{0.2cm}
This concludes the proof.
\hfill 
\IEEEQED
\end{appendices}
\vspace{0.3cm}
\def\baselinestretch{.88}
\bibliographystyle{IEEEbib}
\bibliography{main}

\begin{thebibliography}{10}

\bibitem{Omid2019}
O.~{Semiari}, W.~{Saad}, M.~{Bennis}, and M.~{Debbah},
\newblock ``Integrated millimeter wave and sub-6 ghz wireless networks: A
  roadmap for joint mobile broadband and ultra-reliable low-latency
  communications,''
\newblock {\em IEEE Wireless Communications}, vol. 26, no. 2, pp. 109--115,
  2019.

\bibitem{3gpp_st}
3GPP,
\newblock ``Nr; physical channels and modulation, technical specification,''
\newblock {\em 3rd Generation Partnership Project}, vol. Rel 15, 2018.

\bibitem{Mao2020}
J.~{Mao}, L.~{Zhang}, P.~{Xiao}, and K.~{Nikitopoulos},
\newblock ``Interference analysis and power allocation in the presence of mixed
  numerologies,''
\newblock {\em IEEE Transactions on Wireless Communications}, vol. 19, no. 8,
  pp. 5188--5203, 2020.

\bibitem{Kihero2019}
A.~B. {Kihero}, M.~S.~J. {Solaija}, and H.~{Arslan},
\newblock ``Inter-numerology interference for beyond 5g,''
\newblock {\em IEEE Access}, vol. 7, pp. 146512--146523, 2019.

\bibitem{syamfOFDM2020}
K.~S. {Chandran} and C.~K. {Ali},
\newblock ``Filtered-ofdm with index modulation for mixed numerology
  transmissions,''
\newblock in {\em 2020 6th International Conference on Advanced Computing and
  Communication Systems (ICACCS)}, 2020, pp. 306--310.

\bibitem{akhtar2018}
A.~{Akhtar} and H.~{Arslan},
\newblock ``Downlink resource allocation and packet scheduling in
  multi-numerology wireless systems,''
\newblock in {\em 2018 IEEE Wireless Communications and Networking Conference
  Workshops (WCNCW)}, 2018, pp. 362--367.

\bibitem{Marij2018}
L.~{Marijanović}, S.~{Schwarz}, and M.~{Rupp},
\newblock ``Optimal resource allocation with flexible numerology,''
\newblock in {\em 2018 IEEE International Conference on Communication Systems
  (ICCS)}, 2018, pp. 136--141.

\bibitem{Liu2020}
X.~{Liu}, L.~{Zhang}, J.~{Xiong}, X.~{Zhang}, L.~{Zhou}, and J.~{Wei},
\newblock ``Peak-to-average power ratio analysis for ofdm-based
  mixed-numerology transmissions,''
\newblock {\em IEEE Transactions on Vehicular Technology}, vol. 69, no. 2, pp.
  1802--1812, 2020.

\bibitem{cheng2019}
X.~{Cheng}, R.~{Zayani}, H.~{Shaiek}, and D.~{Roviras},
\newblock ``Inter-numerology interference analysis and cancellation for massive
  mimo-ofdm downlink systems,''
\newblock {\em IEEE Access}, vol. 7, pp. 177164--177176, 2019.

\bibitem{abu2018}
A.~{T.Abusabah} and H.~{Arslan},
\newblock ``Noma for multinumerology ofdm systems,''
\newblock {\em Wireless Communications and Mobile Computing}, vol. 2018, pp.
  1--9, 05 2018.

\bibitem{Bag2019}
T.~{Bag}, S.~{Garg}, Z.~{Shaik}, and A.~{Mitschele-Thiel},
\newblock ``Multi-numerology based resource allocation for reducing average
  scheduling latencies for 5g nr wireless networks,''
\newblock in {\em 2019 European Conference on Networks and Communications
  (EuCNC)}, 2019, pp. 597--602.

\bibitem{gok2020}
S.~{Gökceli}, T.~{Levanen}, J.~{Yli-Kaakinen}, T.~{Riihonen}, M.~{Renfors},
  and M.~{Valkama},
\newblock ``Papr reduction with mixed-numerology ofdm,''
\newblock {\em IEEE Wireless Communications Letters}, vol. 9, no. 1, pp.
  21--25, 2020.

\bibitem{cheng2019UL}
X.~{Cheng}, R.~{Zayani}, H.~{Shaiek}, and D.~{Roviras},
\newblock ``Analysis and cancellation of mixed-numerologies interference for
  massive mimo-ofdm ul,''
\newblock {\em IEEE Wireless Communications Letters}, vol. 9, no. 4, pp.
  470--474, 2020.

\bibitem{yang2020}
B.~{Yang}, L.~{Zhang}, O.~{Onireti}, P.~{Xiao}, M.~A. {Imran}, and
  R.~{Tafazolli},
\newblock ``Mixed-numerology signals transmission and interference cancellation
  for radio access network slicing,''
\newblock {\em IEEE Transactions on Wireless Communications}, vol. 19, no. 8,
  pp. 5132--5147, 2020.

\bibitem{23GPP38}
\text{3GPP TS 38.300, v.15.2.0},
\newblock ``{NR; NR and NG-RAN Overall Description; Stage 2},''
\newblock June 2018.

\bibitem{boyd_vandenberghe_2004}
L.~{Vandenberghe} S.~{Boyd},
\newblock {\em Convex Optimization},
\newblock Cambridge University Press, 2004.

\bibitem{poutageconvex}
G.~{Caire}, G.~{Taricco}, and E.~{Biglieri},
\newblock ``Optimum power control over fading channels,''
\newblock {\em IEEE Transactions on Information Theory}, vol. 45, no. 5, pp.
  1468--1489, 1999.

\end{thebibliography}
\end{document}